\input{aipcheck}

\documentclass[draft,]{aipproc}

\layoutstyle{6x9}

\begin{document}

\title{Abrikosov Gluon Vortices in Color Superconductors}


\author{Efrain J. Ferrer}{
  address={Department of Physics\\
The University of Texas at El Paso\\
El Paso,
TX 79968,
USA}
}

\begin{abstract}
In this talk I will discuss how the in-medium magnetic field can influence the gluon dynamics in a three-flavor color superconductor. It will be shown how at field strengths comparable to the charged gluon Meissner mass a new phase can be realized, giving rise to Abrikosov's vortices of charged gluons. In that phase, the inhomogeneous gluon condensate anti-screens the magnetic field due to the anomalous magnetic moment of these spin-1 particles. This paramagnetic effect can be of interest for astrophysics, since due to the gluon vortex antiscreening mechanism, compact stars with color superconducting cores could have larger magnetic fields than neutron stars made up entirely of nuclear matter. I will also discuss a second gluon condensation phenomenon connected to the Meissner instability attained at moderate densities by two-flavor color superconductors. In this situation, an inhomogeneous condensate of charged gluons emerges to remove the chromomagnetic instability created by the pairing mismatch, and as a consequence, the charged gluonic currents induce a magnetic field. Finally, I will point out a possible relation between glitches in neutron stars and the existence of the gluon vortices.
\end{abstract}

\maketitle


\section{Introduction}

  The high baryon densities of neutron-star cores makes this natural environment a special place for the possible realization of deconfined quark matter. If this quark phase is reached at relatively low temperatures, the attractive strong interaction in the antitriplet channel produces an instability in the system's perturbative ground state (i.e. the famous Cooper instability) that results in the formation of diquark Cooper pairs, and consequently settling the phenomenon of color superconductivity \cite{Bailin-Love}. On the other hand, compact stars also develop very intense magnetic fields, that in the surface of the so called magnetars can reach values perhaps as high as $10^{16} G$ \cite{magnetars}. Taking into account that the magnetic flux is conserved in such a medium, the magnetic field can even have larger values in the inner core when increasing the density. Theoretical estimates based on the virial theorem give for these gravitational bound objects a maximum value of $10^{19} G$ \cite{virial}. Then, it is imperative to study magnetism in color superconductivity.

  Similarly to what occurs with the electroweak phase transition, the electromagnetic field in spin-zero color superconductivity (CS) is redefined once the ground state is restructured by the condensation of the Cooper pairs. The new long-range electromagnetic potential in the color superconductor, $\widetilde{A}_{\mu}=\cos{\theta}\,A_{\mu}-\sin{\theta}\,G^{8}_{\mu}$, is given by the linear combination of the conventional electromagnetic potential, $A_\mu$, and that of the $8^{th}$ gluon, $G^8_\mu$, \cite{alf-raj-wil}. The mixing angle $\theta$ is a function of the strong $g$ and electromagnetic $e$ coupling constants, and depends on the nature of the color superconducting phase. Thus, $\widetilde{A}_{\mu}$ is a massless field that becomes the in-medium physical mode (it is called the rotated electromagnetic field). The Cooper pairs, although are charged with respect to the conventional electromagnetism, are neutral with respect to the rotated one. As a consequence, there is no Meissner effect for a rotated magnetic field in the spin-zero CS. Another peculiarity of this medium is the existence of several scales that can be associated with different magnetic effects. At first glance, one might think that the magnetic field strength needed to produce any physical effect in a CS medium with baryon density of order $\mu \sim 400-500$ MeV would have to be comparable to that scale, meaning, close or larger than the expected maximum value in the inner core of a neutron star. Nevertheless, as shown in a series of papers \cite{MCFL}-\cite{Oscillations}, for magnetic fields of strengths comparable to the two other CS scales (i.e. the dynamically generated gap, $\Delta$, and the gluon Meissner mass, $m_{g}$) new physical effects can manifest. Assuming that at
sufficiently high $\mu$, the running strong coupling $g$ becomes
$g(\mu)\ll 1$, the hierarchy of the scales is $\Delta\ll
m_{g}\ll \mu$.

As shown in Ref \cite{MCFL}, the
color-superconducting properties of a three-flavor
massless quark system are substantially
affected by the penetrating $\widetilde{H}$ field and as a
consequence, a new phase, called Magnetic Color Flavor Locked (MCFL)
phase, takes place. In the MCFL phase the gaps that receive contributions from pairs of charged quarks get reinforced at very strong fields producing a sizable splitting as compared with the gaps that only get contribution from pairs of neutral quarks. As the field is decreased, the gaps become oscillating functions of the magnetic field, a phenomenon which is associated to the known Haas-van Alphen oscillations appearing in magnetized systems \cite{Oscillations}.

Although the symmetry breaking patterns of the MCFL and CFL phases
are different \cite{MCFL}, the two phases are hardly distinguishable at weak
magnetic fields. Now, in the CFL phase the symmetry breaking leaves nine Goldstone bosons: a singlet
associated to the breaking of the baryonic symmetry $U(1)_B$, and
an octet associated to the axial $SU(3)_A$ group.
Once a magnetic field is switched on, the difference between the
electric charge of the $u$ quark and that of the $d$ and $s$
quarks reduces the original flavor symmetry of the theory and
consequently also the symmetry group remaining after the diquark
condensate is formed. Then, the breaking pattern only leaves five
Goldstone bosons \cite{PCFL}. Three of them correspond to the breaking
of $SU(2)_A$, one to the breaking of $U(1)^{(1)}_A$ (not to be confused with the usual
anomaly $U(1)_{A}$), which is related to the current which is an
anomaly-free linear combination of $s$, $d$, and $u$ axial
currents \cite{miransky-shovkovy-02}, and one to
the breaking of $U(1)_B$. Thus, an applied magnetic field reduces
the number of Goldstone bosons in the superconducting phase, from
nine to five.

For a meson to be stable, its mass
should be less than twice the gap, otherwise it will decay into a
particle-antiparticle pair. That means that the threshold field
for the effective $CFL \rightarrow MCFL$ symmetry transmutation is
given by \cite{PCFL}
\begin{equation}\label{Threshold-value}
\widetilde{e}\widetilde{H}_{MCFL}=\frac{4}{v_{\bot}^2}\Delta_{CFL}^2\simeq
12\Delta_{CFL}^2.
\end{equation}
 For $\Delta_{CFL} \sim 15 MeV$ one gets ${\tilde e}
{\tilde H}_{MCFL} \sim 10^{16}G$. Hence, the MCFL phase becomes relevant already at moderately large fields, and consequently the gap oscillating effect previously mentioned \cite{Oscillations} starts to manifest.

At the  threshold field, the
charged mesons get masses that are twice the gap and therefore decouple from the low-energy theory. When this
decoupling occurs, the five neutral Goldstone bosons (including the
one associated to the baryon symmetry breaking) that characterize
the MCFL phase will drive the low-energy physics of the system.
Notice, moreover, that the MCFL phase is not just characterized by a smaller number of
Goldstone fields, but by the fact that all these bosons are
neutral with respect to the rotated electric charge. Hence, no
charged low-energy excitation can be produced in the MCFL phase.
This effect can be relevant for the cooling and transport properties of compact stars, since they are
determined by the particles with the lowest energy.

More recently, we have found that the magnetic field can also
influence the gluon dynamics \cite{Vortex}. At field strengths
comparable to the charged gluon Meissner mass $m_M$, a new phase can be
realized giving rise to a vortex arrangement of
$\widetilde{Q}$-charged gluons \cite{Vortex}. The gluon vortices
anti-screens the magnetic field due to the anomalous magnetic
moment of these spin-1 particles. Because of the anti-screening,
this condensate does not give a mass to the $\widetilde{Q}$
photon, but instead amplifies the applied rotated magnetic field.
This means that at such applied fields the CS behaves as a
paramagnet. Thus at $\widetilde{H} \geq m_M^2$ the so called  paramagnetic CFL phase (PCFL) takes place
\cite{PCFL}. This last effect is also of interest for
astrophysics. Compact stars with color superconducting cores could
have larger magnetic fields than neutron stars made up entirely of
nuclear matter, thanks to the gluon vortex antiscreening
mechanism. In the next Section we discuss some peculiarities of this PCFL phase.

\section{Gluon Vortices and Magnetic Antiscreening}

Even though
gluons are neutral with respect to the conventional
electromagnetism, in the $CFL$ phase some of them are charged with respect
to the rotated field. Specifically, the charged gluons $G_{\mu}^{\pm}\equiv\frac{1}{\sqrt{2}}[G_{\mu}^{4}\mp
iG_{\mu}^{5}]$ and
$I_{\mu}^{\pm}\equiv\frac{1}{\sqrt{2}}[G_{\mu}^{6}\mp
iG_{\mu}^{7}]$  acquire ($\pm 1$)  $\widetilde{Q}$ charges given in units of $\widetilde{e} = e \cos{\theta}$. As shown in Ref. \cite{Vortex}, the external rotated magnetic field $\widetilde{H}$
produces an instability when its
strength becomes larger than the Meissner mass square, $m_{M}^2$, of the
$\widetilde{Q}$-charged gluons. To remove the tachyonic
mode, the ground state is restructured through the formation of a
gluon condensate of amplitude $G$, as well as an induced magnetic
field $\widetilde{B}$ that originates due to the backreaction of the
inhomogeneous $G$ condensate on the rotated electromagnetic field. The condensate
solutions can be found by minimizing with respect to $G$ and
$\widetilde{B}$ the Gibbs free-energy density
$\mathcal{G}_{c}=\mathcal{F}-\widetilde{H}\widetilde{B}$,
($\mathcal{F}$ is the free energy density)
\begin{eqnarray}
\label{Gibbs} \mathcal{G}_{c} =\mathcal{F}_{n0}
-2G^{\dag}\widetilde{\Pi}^{2}
G-2(2\widetilde{e}\widetilde{B}-m_{M}^{2})|G|^{2}+2g^{2}|G|^{4} +
\frac{1}{2}\widetilde{B}^{2}-\widetilde{H}\widetilde{B}
\end{eqnarray}
In (\ref{Gibbs}) $\mathcal{F}_{n0}$ is the system free energy
density in the normal-CFL phase ($G=0$) at zero applied field. Using
(\ref{Gibbs}) the minimum equations at $\widetilde{H}\sim
\widetilde{H}_{C}$ for the condensate $G$ and induced field
$\widetilde{B}$ respectively are
\begin{equation}
\label{G-Eq} -\widetilde{\Pi}^{2}
G-(2\widetilde{e}\widetilde{B}-m_{M}^{2})G=0,
\end{equation}
\begin{equation}
\label{B-Eq} 2\widetilde{e} |G|^{2}-\widetilde{B}+\widetilde{H}=0
\end{equation}
Identifying $G$ with the complex order parameter, Eqs.
(\ref{G-Eq})-(\ref{B-Eq}) become analogous to the Ginzburg-Landau
equations for a conventional superconductor except for the
$\widetilde{B}$ contribution in the second term in (\ref{G-Eq}) and
the sign of the first term in (\ref{B-Eq}). The origin of both terms
can be traced back to the anomalous magnetic moment term
$4\widetilde{e}\widetilde{B}|G|^{2}$ in the Gibbs free energy
density (\ref{Gibbs}). Notice that because of the different sign in
the first term of (\ref{B-Eq}), contrary to what occurs in
conventional superconductivity, the resultant field $\widetilde{B}$
is stronger than the applied field $\widetilde{H}$. Thus, when a
gluon condensate develops, the magnetic field will be antiscreened
and the color superconductor will behave as a paramagnet. The
antiscreening of a magnetic field has been also found in the context
of the electroweak theory for magnetic fields $H \geq
M_{W}^{2}/e\sim 10^{24} G$ \cite{Olesen}.

The explicit solution of (\ref{G-Eq})-(\ref{B-Eq}) can be found
following Abrikosov's approach \cite{Abrikosov} to type II metal
superconductivity for the limit situation when the applied field is
near the critical value $H_{c2}$. In our case we find \cite{Vortex}
that the ground state is restructured, forming a vortex state
characterized by the condensation of charged gluons and the creation
of magnetic flux tubes. In the vortex state the magnetic field
outside the flux tubes is equal to the applied one, while inside the
tubes its strength increases by an amount that depends on the
amplitude of the gluon condensate. This non-linear paramagnetic
behavior of the color superconductor can be relevant to boosting the
star's magnetic field without appealing to a magnetohydrodynamic
dynamo mechanism, and it can be important for the physics of magnetars \cite{W-Condensation}.

It is also worth to mention that while in the phenomenon found in Ref. \cite{Vortex}, and discussed above, the condensation of gluons was the response
of the color superconductor to an externally applied and
sufficiently strong magnetic field, that is, a
\textit{magnetic-field-induced phenomenon}, in a different context a gluon condensate with similar characteristics can be induced at zero applied field. The
condensation phenomenon in that case is connected to a Meissner instability
triggered at moderate densities by a pairing stress associated with
the neutrality and $\beta$-equilibrium constraints. Such instability
occurs in the gapped 2SC color superconductor at moderate densities
when the effective Meissner mass of the charged gluons becomes
tachyonic \cite{Huang}. As shown in Ref. \cite{W-Condensation}, in this situation an
inhomogeneous condensate of charged gluons emerges to remove the
chromomagnetic instability created by the pairing mismatch. Similarly to the previous case, the
currents associated to the inhomogeneous condensate of charged
gluons induce a non-zero rotated magnetic field in the system.
The spontaneous induction of a magnetic field in CS systems
that have pairings with mismatched Fermi surfaces is a new kind of
phenomenon that can serve to generate magnetic fields in stellar
compact objects.

\section{Astrophysical Connotations}

In addition to the already mentioned possible astrophysical implications of the different magnetized phases,
I want to call attention about another potential astrophysical consequence of CS in neutron stars, this is the relation between the existence of gluon vortices and glitches in pulsars \cite{glitches-pulsars}.
It is generally accepted that the interior region of neutron stars, due to their rotation and to the existence of strong magnetic fields, contains
superfluid vortices of neutron Cooper pairs and superconducting vortices of proton Cooper pairs \cite{N-P-pairs}. Interaction between the superfluid vortices and the superconducting flux tubes has been pointed out as the possible cause of glitches \cite{glitches}. In the inner core, if CS is realized, the rotating star will create superfluid vortices of the diquark condensate \cite{Baym}; and if the field is strong enough to produce gluon vortices \cite{Vortex}, superconducting flux tubes associated to these vortices may exist. Thus, by the same mechanism which takes place in the outer core we can have that the glitches can get contribution from the quark matter in the core.


\begin{theacknowledgments}
  This work was supported in part by the Office of Nuclear Physics of the Department of Energy under contract DE-FG02-09ER41599.
\end{theacknowledgments}

\bibliographystyle{aipproc}



\bibliographystyle{aipproc}   

\bibliography{sample}

\IfFileExists{\jobname.bbl}{}
 {\typeout{}
  \typeout{******************************************}
  \typeout{** Please run "bibtex \jobname" to optain}
  \typeout{** the bibliography and then re-run LaTeX}
  \typeout{** twice to fix the references!}
  \typeout{******************************************}
  \typeout{}
 }

\end{document}